\documentclass{article}

\input epsf
\begin{document}

\begin{center}
{\bf \Large Nonlinear waves and related nonintegrable and
integrable systems}

\bigskip

{\large N.~A.~Kostov}
\bigskip

{\sl Institute of Electronics\\
Bulgarian Academy of Sciences\\
Blvd. Tsarigradsko shosse 72, Sofia 1784,
\\Bulgaria}
\medskip

E-mail: nakostov@ie.bas.bg

\end{center}

\begin{abstract}
Spectral method related to Lam\'{e} equation with finite-gap
potential is used to study the optical cascading equations. These
equations are known not to be integrable by inverse scattering
method. Due to "partial integrability" two-gap solutions are
obtained in terms of products of elliptic functions and are
classified in five different families related to eigenvalues of
appropriate spectral problem. In special cases, when periodic
solutions reduce to localized solitary waves, previously known
phase-locked solutions are recovered, and additional one solution
is obtained. For vector nonlinear Schr\"{o}dinger equation $n=3$
we present exact solutions in a form of multicomponent cnoidal
waves.
\end{abstract}

\section{Introduction}

The aim of the  present paper is devoted to the rather old but
still open problem how to construct exact periodic solutions of
integrable and nonintegrable soliton systems. This problem is
important from the physical point of view to study nonlinear
waves. These solutions are expressed in terms of Hermite and
Lam\'e polynomials. Results are presented both for integrable and
nonintegrable dynamical systems. As examples optical cascading
equations and  vector nonlinear Schr\"{o}dinger equation are
considered.

\section{Optical cascading equations and phase-locked solutions}

We consider $\chi^{(2)} :\chi^{(2)}$ cascading equations in the
normalized form
\begin{eqnarray} \label{OCE}
&& i a_{1t}-\frac{r}{2} a_{1xx}+ a_{1}^{*}a_{2}=0,\nonumber \\
&& i a_{2t}-\beta a_{2}- i\delta a_{2x}-\frac{\alpha}{2}
a_{2xx}+a_{1}^2=0,
\end{eqnarray}
where $a_{1}$ and $a_{2}$ are the normalized complex envelopes of
FW and SHW, respectively, $t$ is the normalized distance along
the wave guide, and $x$ is the normalized transverse coordinate.
The real constant $\alpha$ is given by minus ratio of the wave
numbers of the FW and SHW, the quantity $\beta$ corresponds to
the normalized wave number mismatch and the parameter $\delta$
corresponds to normalized walkoff coefficient, $r=\pm 1$.

   We seek solution of (\ref{OCE}) in the following form
\begin{eqnarray}
&& a_{1}=q_{1}(\xi) e^{i\phi_{1}(\xi)} e^{i(k_1 t-\omega_{1}x)},
\\
&& a_{2}=q_{2}(\xi) e^{i\phi_{2}(\xi)} e^{2i(k_1 t-\omega_{1}x)},
\end{eqnarray}
with $\xi=\omega x -v t$ and we define $\phi=\phi_{2}-2\phi_{1}$.
We will consider phase-locked solutions ($\phi=k\pi$, $k\in {\bf
Z})$, then the system (\ref{OCE}) is reduced to two ordinary
differential equations,
\begin{eqnarray}
&& q_{1\xi\xi}+ A_{0} q_{1} + B_{0} q_{1} q_{2} =0, \label{ROCE1}\\
&& q_{2\xi\xi}+ C_{0} q_{2} + D_{0} q_{2}^{2} =0,   \label{ROCE2}
\end{eqnarray}
where we have
\begin{eqnarray}
A_{0}=A^{2}-D,\, B_{0}=-\frac{2\epsilon}{r\omega^{2}},\,
C_{0}=B^{2}-C, \,D_{0}=-\frac{2\epsilon}{\alpha\omega^{2}}, \,
\epsilon^2=1,
\end{eqnarray}
and
\begin{eqnarray}
&& A=\frac{r\omega\omega_{1}-v}{r\omega^{2}},\quad
B=\frac{2\alpha\omega\omega_{1}-v-\delta\omega}{\alpha\omega^{2}}
\nonumber\\
&&C=2\frac{2\alpha\omega_{1}^{2}-2k_{1}-2\delta\omega_{1}-\beta}
{\alpha\omega^{2}},\quad
D=\frac{r\omega_{1}^{2}-2k_{1}}{r\omega^{2}}.
\end{eqnarray}
Introducing new variable
\begin{eqnarray}
q_{1}^{2}=\frac{4 F}{B_{0}D_{0}}, \quad
F=\lambda^{2}-3\wp\lambda+9\wp^{2}-\frac{9}{4}g_{2}
\end{eqnarray}
where $F$ is Hermite polynomial \cite{ww86}, $g_{2}, g_{3}$ are
elliptic invariants defined in \cite{ww86}. $\wp=\wp(\xi+\omega')$
is Weierstrass function shifted by half period $\omega'$ is
related to $\mbox{sn}$ Jacobian elliptic function with modulus $k$
\begin{eqnarray}
\wp(\xi+\omega';g_2,g_3)=\alpha^2 k^2\mbox{sn}^2(\alpha
\xi,k)-(1+k^2),
\end{eqnarray}
where $\alpha=\sqrt{e_1 - e_3}$ and $e_{i},\,$$i=1,2,3,\,e_3\leq
e_2\leq e_1$ are the real roots of the cubic equation
\begin{equation}
4\lambda^3-g_2\lambda-g_3=0.
\end{equation}
Using wave height $\alpha$ and modulus
$k=\sqrt{\frac{e_2-e_3}{e_1-e_2}}$ we have the following relations
\begin{eqnarray}
&&e_1=\frac{1}{3}(2-k^2)\alpha^2,\quad
e_2=\frac{1}{3}(2k^2-1)\alpha^2,\quad
e_3=-\frac{1}{3}(1+k^2)\alpha^2,\nonumber \\
&& g_2=-4(e_1 e_2 + e_1 e_3 + e_2
e_3)=\frac{4}{3}\alpha^2(1-k^2+k^4),\nonumber \\
&& g_3=4e_1 e_2 e_3=\frac{4}{27}\alpha^6(k^2+1)(2-k^2)(1-2k^2).
\end{eqnarray}
Inserting this expression in (\ref{ROCE1}) we have the following
nonlinear differential equation with spectral parameter
$\lambda=-C_{0}/2$
\begin{eqnarray} \label{Feq}
\frac{1}{2} F F_{\xi\xi} -\frac{1}{4} F_{\xi}^{2}-
(u(\xi)+\lambda) F^{2} +\frac{1}{4} R(\lambda)=0,
\end{eqnarray}
with eigenvalue equations
\begin{eqnarray}
&&R(\lambda)=4\lambda^{5}-21\lambda^{3} g_{2} +27\lambda g_{2}^2 +
27 \lambda^2 g_{3}-81 g_{2} g_{3}=0,\nonumber \\ &&u(\xi)=-(B_{0}
q_{2}+\lambda+A_{0}) =6\wp(\xi+\omega'),
\end{eqnarray}

or in factorized form
\begin{eqnarray}
&&R(\lambda)=4\prod_{i=1}^{5} (\lambda-\lambda_{i})=0, \quad
\lambda_{1}=-\sqrt{3g_{2}},\quad \lambda_{2}=3 e_{3} \nonumber \\
&&\lambda_{3}=3 e_{2},\quad \lambda_{4}= 3 e_{1}, \quad
\lambda_{5}=\sqrt{3g_{2}}. \end{eqnarray}
 It is well known that
equation (\ref{Feq}) is reduced to linear periodic spectral
problem of one dimensional Schr\"{o}dinger equation with two gap
potential $u(x)=6\wp(\xi+\omega')$ and with five normalized
eigenfunctions $q_{1}^{(i)}$, $(i)=1,\ldots 5$:
\begin{eqnarray}
\frac{d^2 q_{1}^{(i)}}{d^2\xi^2}-u(\xi) q^{(i)}_{1}=\lambda_{i}
q^{(i)}_{1}, \quad (i)=1,\ldots ,5 .
\end{eqnarray}
Under these conditions the second equation (\ref{ROCE2}) is
automatically satisfied. Second equation can be considered as
"self-consistent" equation for potential $u(\xi)$. Finally the
five spectral families of periodic solutions can be written in the
following Table \ref{tab:1}
\begin{center}
\begin{table}[h]
\caption{Five spectral families of periodic solutions
\label{tab:1}}
\begin{tabular}{|l|l|l|l|} \hline
(I).  & $q_{1}= \frac{6}{\sqrt{B_{0}D_{0}}}\alpha^2 k^2\,
\mbox{E}_{2}^{(u-)}$ & $q_{2}=-\frac{1}{B_{0}}(u(\xi)+\frac{3
g_{2}} {\lambda_{1}}-2\lambda_{1})$ &(i)=1\\ \hline (II) & $q_{1}=
\frac{6}{\sqrt{B_{0}D_{0}}}\alpha^2 k\,\mbox{E}^{(cd)}_{2}$ & $
q_{2}=-\frac{1}{B_{0}}
(u(\xi)+\frac{3 g_{2}}{\lambda_{2}}-2\lambda_{2}) $ & (i)=2\\
\hline (III) & $q_{1}= \frac{6}{\sqrt{B_{0}D_{0}}}\alpha^2 k
\,\mbox{E}^{(sd)}_{2}$ & $ q_{2}=-\frac{1}{B_{0}}(u(\xi)+\frac{3
g_{2}}{\lambda_{3}}-2\lambda_{3})$ & (i)=3  \\ \hline (IV) &
$q_{1}= \frac{6}{\sqrt{B_{0}D_{0}}}\alpha^2 k^2\,
\mbox{E}^{(sc)}_{2}$ &
$ q_{2}=-\frac{1}{B_{0}}(u(\xi)+\frac{3 g_{2}}{\lambda_{4}}-2\lambda_{4}) $  & (i)=4\\
\hline (V) & $q_{1}= \frac{6}{\sqrt{B_{0}D_{0}}}\alpha^2 k^2\,
\mbox{E}_{2}^{(u+)}$ & $ q_{2}=-\frac{1}{B_{0}}(u(\xi)+\frac{3
g_{2}}{\lambda_{5}}-2\lambda_{5}) $ & (i)=5 \\
\hline \end{tabular}
\end{table}
\end{center}
where
\begin{eqnarray}
&&\mbox{E}^{(sc)}_{2}=\mbox{sn}(\alpha \xi,k)\mbox{cn}(\alpha \xi,k),\nonumber\\
&&\mbox{E}^{(sd)}_{2}=\mbox{sn}(\alpha \xi,k)\mbox{dn}(\alpha \xi,k),\nonumber\\
&&\mbox{E}^{(cd)}_{2}=\mbox{cn}(\alpha \xi,k)\mbox{dn}(\alpha \xi,k),\nonumber\\
&&\mbox{E}_{2}^{(u\pm)}=\mbox{sn}^2(\alpha \xi,k)-
\frac{1+k^2\pm\sqrt{1-k^2+k^4}}{3k^2},    \nonumber\\
\end{eqnarray}
are normalized two-gap Lam\'{e} functions \cite{ww86},
$\mbox{cn}$, $\mbox{dn}$ are Jacobian elliptic functions and
potential $u(\xi)$ have the form
\begin{eqnarray}
 u(\xi)=6 \alpha^2 k^2\mbox{sn}^{2}(\alpha \xi,k)-2(1+k^2)\alpha^2.
\end{eqnarray}
If, however, we restrict ourselves to using solitary waves that
corresponds to limit $k\rightarrow 1$, for FW $q_{1}$ we have the
following forms (case I):
$\,\mbox{sech}^{2}(\alpha\xi)-\frac{2}{3}$; case (II) and (V):
$\,\mbox{sech}^{2}(\alpha\xi)$; case (III) and (IV):
$\,\mbox{tanh}(\alpha\xi)\mbox{sech}(\alpha\xi)$, the red, white
and blue solitary waves respectively. These wavetrains, so called
order two solitary waves are well known phenomena only in the case
of coupled nonlinear Schr\"{o}dinger equations \cite{hioe98},
\cite{aa99}.

The theory of optical cascading in materials with a pure quadratic
(or $\chi^{(2)}$) nonlinearity became the subject of many
theoretical and experimental investigations. In the case of
$\chi^{(2)}$ materials it has been possible to produce solitary
waves and periodic solutions as background trough $\chi^{(2)}
:\chi^{(2)}$ cascading. This effect occurs in parametrically
coupled fields with quadratic nonlinearities and interacting
fundamental (FW) and second harmonic (SHW) waves. Several
particular wave solutions of the system describing this phenomenon
have been obtained, e.g. with the aid of the Hamiltonian formalism
\cite{ft95}, direct substitution \cite{p98,pt99,pp02}, and Lie
group analysis \cite{lwm98}. General families of periodic waves
are reported recently in \cite{kvt03, kezvt04}.

\section{Vector nonlinear Schr\"odinger equation}

We consider the system of coupled nonlinear Schr\"odinger
equations
\begin{eqnarray}
i \frac{\partial}{\partial t} Q_{j}+s \frac{\partial^{2}}{\partial
x^2}Q _{j}+\sigma\left(\sum_{k=1}^{n}|Q_{k}|^{2}\right) Q_{j}=0,
\quad j=1,\dots ,n, \label{VNLSE}
\end{eqnarray}
where $s=\pm 1$, $\sigma=\pm 1$.  These equations are important
for a number of physical applications. For example, for
photorefractive media with a drift mechanism of nonlinear
response, a good approximation describing the propagation of $n$
self-trapped mutually incoherent wave packets is the set of
equations for a Kerr-type nonlinearity \cite{kpsv98}
\begin{eqnarray}
i \frac{\partial}{\partial {z'}}\tilde{Q}_{j}+\frac{1}{2}
\frac{\partial^{2}}{\partial
{x'}^{2}}\tilde{Q}_{j}+\alpha\delta\eta \tilde{Q}_{j}=0, \quad
j=1,\ldots ,n , \label{photorefEq}
\end{eqnarray}
where $\tilde{Q}_{j}$ denotes the $j$th component of the beam,
$\alpha$ is a coefficient representing the strength of
nonlinearity, $z'$ and $x'$ are the coordinate along the direction
of propagation and transverse coordinate respectively. The change
in refractive index profile $\eta$ created by all the incoherent
components in the light beam is defined by
\begin{eqnarray}
\delta\eta  = \sum_{k=1}^{n}|\tilde{Q}_{k}|^{2}. \label{index}
\end{eqnarray}
Inserting (\ref{index}) in (\ref{photorefEq}) and renormalising
the variables as $\tilde{Q}_{j}=Q_{j}/\sqrt{2\alpha}$,
$z'=2t,x'=x$ we obtain the vector nonlinear Schr\"{o}dinger
equation (\ref{VNLSE}). Stability, localization, and soliton
asymptotics of multicomponent photorefractive cnoidal waves are
discussed in \cite{psv99}. New solutions are presented next for
the case $n=3,4$.

We seek solution of (\ref{VNLSE}) in the following form
\cite{eek00})
\begin{equation}
\label{ansatz}Q_{j}=q_{j}(z)\,\mbox{e}^{i\Theta_{j}},\quad
j=1,\ldots n,
\end{equation}
where $z=x-ct$, $\Theta_{j}=\Theta_{j}(z,t)$, with
$q_{j},\Theta_{j}$ real. Substituting (\ref{ansatz}) into
(\ref{VNLSE}) and separating real and imaginary parts by supposing
that the functions $\Theta_{j}, j=1,\ldots n$ behave as
\begin{eqnarray*}
\Theta_{j}= \frac{1}{2}s c x+(a_{j}-\frac{1}{4}sc^2)t-s\,{\mathcal
C}_{j} \int_{0}^{z}\frac{\mathrm{d} z'}{q_{j}(z')^2} +\Theta_{j0},
\end{eqnarray*}
we obtain the system  $(\sigma= s =\pm 1)$
\begin{eqnarray}
&&\frac{d^2}{dz^2}q_{j}+ \left(\sum_{k=1}^{n}
\frac{\sigma}{s}q_{k}^{2}-\frac{a_{j}}{s}\right) q_{j} -
\frac{{\mathcal C}_j^2}{q_{j}^3}=0, \label{systemn} \quad
k,j=1,\ldots n  ,
\end{eqnarray}
where $C_j,$ $j=1,\ldots n$ are free parameters and $\Theta_{j0}$
are constants. These equations describe the integrable case of
motion of a particle in a quartic potential perturbed with inverse
squared potential, which is separable in ellipsoidal coordinates.
The solutions of the system (\ref{systemn}) are then given as
\begin{equation}
q_{i}^2(z)=2\frac{{\mathcal  F}(z,a_{i}-\Delta)} {\prod_{k\neq
i}^{n} (a_{i}-a_{k})} , \, i=1, \ldots ,n , \label{answern1}
\end{equation}
where ${\mathcal F}(z,\lambda)$ is Hermite polynomial associated
with Lam\'{e} potential. The final formula for the solutions of
the system (\ref{VNLSE}) then reads
\begin{eqnarray}
Q_{i}(x,t)=\sqrt{2\frac{{\mathcal  F}(z,a_{i}-\Delta)}
{\prod_{k\neq i}^{n} (a_{i}-a_{k})} }\, \mathrm{exp}(\Theta_{i}) ,
\label{answern}
\end{eqnarray}
where
\begin{eqnarray}
\Theta_{j}=\left \{\frac{1}{2}i cx+i(a_{j}-\frac{1}{4}c^2)
t-\frac12\nu(a_{j}-\Delta)\int\limits_{0}^{z} \frac{{\mathrm d}
z'}{{\mathcal F}(z',a_{j}-\Delta)}\right\}, \nonumber
\end{eqnarray}
and $i=1, \ldots ,n$ and we have made use of (\ref{answern1}) and
(\ref{ansatz}). To obtain the special class of periodic solution
of (\ref{systemn}) we introduce the following ansatses
\begin{eqnarray}
q_{i}(\zeta)=\sqrt{A_{i}\wp(\zeta+\omega')+B_{i}}
 , \quad i=1,2,3,\,\mbox{or}\,i=1,\ldots ,4. \label{Hansatz-1}
\end{eqnarray}
As a result we obtain:
\begin{eqnarray}
&&\sum_{k=1}^{m}A_{k}=-2,\quad a_{i}=\sum_{k=1}^{m}B_{k}
-\frac{B_{i}}{A_{i}},\,m=3\,\mbox{or}\,4,
\\ &&-\frac{4{\mathcal C}_{i}^2}{A_{i}^2}=(4\lambda^3-
\lambda g_{2}-g_{3})_{|_{\lambda=-B_{i}/A_{i}}},\quad i=1,\ldots
,3\,\mbox{or}\,4 \end{eqnarray} and using the well known relations
\begin{eqnarray}
\int_{0}^{z}\frac{dz'}{\wp(z')-\wp(\tilde{a}_{j})}=
\frac{1}{\wp'(\tilde{a}_{j})}\left(2z\zeta(\tilde{a}_{j})+
\mbox{ln}\frac{\sigma(z-\tilde{a}_{j})}
{\sigma(z+\tilde{a}_{j})}\right),
\end{eqnarray}
and
\begin{eqnarray}
\wp(z+\omega')-\wp(\tilde{a}_{j})=-
\frac{\sigma(z+\omega'+\tilde{a}_{j})\sigma(z+\omega'-\tilde{a}_{j})}
{\sigma(z+\omega')^{2}\sigma(\tilde{a}_{j})^{2} }.
\end{eqnarray}
We derive the following result
\begin{eqnarray}
 Q_{j}=&&\sqrt{-A_{j}}\frac{\sigma(z+\omega'+\tilde{a}_{j}) }
{\sigma(z+\omega')\sigma(\tilde{a}_{j}) } \times \nonumber \\ &&
\exp\left(\frac{i}{2}cx+i(a_{j}-\frac{1}{4}c^2)t-(z+\omega')
\zeta(\tilde{a}_{j})\right), \label{sigman}
\end{eqnarray}
where
\begin{eqnarray}
&&\sum_{j=1}^{\epsilon_{1}}A_{j}=-2,\quad
a_{j}=\sum_{k=1}^{\epsilon_{1}} B_{k}-\frac{B_{j}}{A_{j}},
\nonumber\\ &&\frac{{\mathcal C}_{j}}{A_{j}}=\frac{i}{2}
\sqrt{4\lambda^3-\lambda g_{2}
-g_{3}}_{|\lambda=-\frac{B_{j}}{A_{j}} } \nonumber\\
&&\wp(\tilde{a}_{j})=-\frac{B_{j}}{A_{j}}=\hat{a}_{j},\quad
j=1\ldots \epsilon_{1},\, \epsilon_{1}=3,4
\end{eqnarray}
One special solution is written by
\begin{eqnarray}
q_{j}=C_{j}\mbox{cn}(\alpha z,k),\qquad j=1,2,3,
\end{eqnarray}
where
\begin{eqnarray}
\alpha^2=\frac{a_{1}}{2k^2-1},\quad \sum_{j=1}^{3} C_{j}^{2}=
2\alpha^2 k^2, \quad a_{1}=a_{2}=a_{3}=a,
\end{eqnarray}
in the limit $k\rightarrow 1$ we obtain soliton solution
\begin{eqnarray}
Q_{j}=\frac{ \sqrt{2a} \epsilon_{j}
\exp\left\{i\left(\frac{1}{2}c(x-x_{0})+(a-\frac{1}{4}
c^{2})t\right) \right\} } {\mbox{ch}(\sqrt{a} (x-x_{0}-ct)) },
\qquad j=1,2,3,
\end{eqnarray}
where we introduce the following notations
\begin{eqnarray}
\sum_{k=1}^{3}|\epsilon_{k}|^2=1, \quad \zeta_{1}=\frac{1}{2}
c+i\sqrt{a} =\xi+i\eta,
\end{eqnarray}
where $x_{0}$ is the position of soliton, $\epsilon_{j}, j=1,2,3$
are the components of polarization vector. One notes that the real
part of $\zeta_{1}$ i.e. $c/2$ gives us the soliton velocity while
the imaginary part of $\zeta_{1}$ i.e. $\sqrt{2a}$ gives the
soliton amplitude and width. Another special solution is written
by
\begin{eqnarray}
q_{j}=C_{j}\mbox{sn}(\alpha z,k),\quad j=1,2 ,\quad
q_{2}=C_{2}\mbox{dn}(\alpha z,k),
\end{eqnarray}
where
\begin{eqnarray}
&&\alpha^2=a_{3}-a_{1},\quad
C_{3}^{2}=2a_{3}-a_{1}-\alpha^2 (1-k^2), \nonumber \\
&&\frac{1}{k^2}\sum_{j=1}^{2}C_{j}^{2}=a_{1}-\alpha^2
(1-k^2),\nonumber
\end{eqnarray}
in the limit $k\rightarrow 1$ we obtain the following soliton
solution
\begin{eqnarray}
Q_{j}&=&\sqrt{a} \epsilon_{j}
\exp\left\{i\left(\frac{1}{2}c(x-x_{0})+(a-\frac{1}{4}
c^{2})t\right) \right\}  \times \nonumber \\ &&
\mbox{th}(\alpha (x-x_{0}-ct)) ,\nonumber \\
Q_{3}&=&\frac{ \sqrt{a+2\alpha^2} \epsilon_{3}
\exp\left\{i\left(\frac{1}{2}c(x-x_{0})+(a+\alpha^2-\frac{1}{4}
c^{2})t\right) \right\} } {\mbox{ch}(\alpha (x-x_{0}-ct))
},\nonumber
\end{eqnarray}
where we introduce the following notations
\begin{eqnarray}
\sum_{j=1}^{2}|\epsilon_{j}|^2=1, \quad |\epsilon_{3}|^2=1,\quad
a_{1}=a_{2}=a=\sum_{j=1}^{2}C_{j}^2, \, a_{3}=a+\alpha^2.
\nonumber
\end{eqnarray}

To obtain the class of periodic solutions of system
(\ref{systemn}) for $n=3,4$ we introduce the following two
ansatses in terms of the Weierstrass function $\wp(\zeta+\omega')$
\begin{eqnarray}
q_{i}(\zeta)=\sqrt{A_{i}\wp(\zeta+\omega')^3+B_{i}\wp(\zeta+\omega')^2+
C_{i}\wp(\zeta+\omega') + D_{i}} ,  \label{Hansatz}
\end{eqnarray}
where $i=1,\ldots 3$. Next for conciseness we denote
$\wp=\wp(\zeta+\omega')$, then the second ansatz have the form
\begin{eqnarray}
q_{i}(\zeta)=\sqrt{A_{i}\wp^4+B_{i}\wp^3+ C_{i}\wp^2 +
D_{i}\wp+E_{i}} , \nonumber\\ i=1,\ldots 4 \label{Hansatz1}
\end{eqnarray} with the constants $A_{i},B_{i},C_{i},D_{i},E_{i}$
defined from the compatibility condition of the ansatz with the
equations of motion (\ref{systemn}). Inserting (\ref{Hansatz}) and
(\ref{Hansatz1}) into Eqs. (\ref{systemn}), using the basic
equations for Weierstrass $\wp$ function \cite{ww86}
\begin{eqnarray}
\left(\frac{d}{d\zeta}\wp(\zeta)\right)^2=4\wp(\zeta)^3-g_{2}\wp(\zeta)-g_{3}
, \quad
\frac{d^2}{d\zeta^2}\wp(\zeta)=6\wp(\zeta)-\frac{g_{2}}{2},
\end{eqnarray}
and equating to zero the coefficients at different powers of $\wp$
we obtain the following algebraic equations for the parameters of
the solutions $A_{i},B_{i},C_{i},D_{i}, i=1,2,3$ for $n=3$
\begin{eqnarray}
&&A_{1}+A_{2}+A_{3}=0,\qquad B_{1}+B_{2}+B_{3}=0, \label{Peq1}\\
&&C_{1}+C_{2}+C_{3}=-12, \quad
C_{i}=\frac{2}{3}\frac{B_{i}^2}{A_{i}}
-\frac{1}{4}A_{i}g_{2}, \label{Peq2}\\
&& a_{i}=\sum_{i=1}^{3} D_{i} -5\frac{B_{i}}{A_{i}}, \quad
D_{i}=\frac{5}{9}\frac{B_{i}^3}{A_{i}^2}-\frac{1}{3}B_{i}g_2-\frac{1}{4}
A_{i}g_{3}\label{Peq3}.
\end{eqnarray}
The analogical algebraic system for $n=4$ is as follows
\begin{eqnarray}
&&A_{1}+A_{2}+A_{3}+A_{4}=0,\qquad B_{1}+B_{2}+B_{3}+B_{4}=0, \label{Pequ1}\\
&&C_{1}+C_{2}+C_{3}+C_{4}=0,\quad D_{1}+D_{2}+D_{3}+D_{4}=-20,\label{Pequ2}\\
&& C_{i}=\frac{3}{5}\frac{B_{i}^2}{A_{i}}
-\frac{3}{10}A_{i}g_{2},\quad
D_{i}=\frac{14}{45}\frac{B_{i}^3}{A_{i}^2}-\frac{53}{180}B_{i}g_{2}-
\frac{2}{9}A_{1}g_{3} \label{Pequ3}\\
&& E_{i}=\frac{49}{225}\frac{B_{i}^4}{A_{i}^3}
-\frac{113}{450}\frac{B_{i}^{2}}{A_{i}}g_{2}-\frac{11}{36}B_{i}g_{3}+
\frac{9}{400}A_{i}g_{2}^2. \label{Pequ4} \\
&&a_{i}=\sum_{i=1}^{4} E_{i} -7\frac{B_{i}}{A_{i}},   \nonumber
\end{eqnarray}
Another result from the algebraic systems is the expression for
constants ${\mathcal C}_{i}$ which parametrize our solutions. For
them we obtain
\[ {\mathcal C}_{i}^{2}=-\frac{\nu(a_{i}-\Delta)^2}{\prod_{k\neq i}
(a_{i}-a_{k})}, \label{Cconst1}
\]
where $i,k=3\,\mbox{or}\, 4$ and parameters $\nu$ are defined by
(for $n=3$)
\begin{eqnarray}
\nu^2=&&\lambda^7-\frac{63}{2}g_{2}\lambda^5+\frac{297}{2}g_{3}\lambda^4
+\frac{4185}{16}g_{2}^2\lambda^3-\nonumber \\
&&\frac{18225}{8}g_{2}g_{3}\lambda^2+\frac{91125}{16}g_{3}^{2}\lambda
-\frac{3375}{16}g_{2}^2\lambda, \label{lame3}
\end{eqnarray}
and (for $n=4$)
\begin{eqnarray}
\nu^2&=&\lambda^9-\frac{231}{2}\lambda^7
g_{2}+\frac{2145}{2}g_{3}\lambda^6+ \frac{63129}{16}\lambda^5
g_{2}^2-\frac{518505}{8}g_{2} g_{3}\lambda^4 \nonumber
\\&&+\left(-\frac{563227}{16}g_{2}^3+\frac{4549125}{16}g_{3}^2\right)
\lambda^3+\frac{991515}{2} g_{3}g_{2}^2\lambda^2+ \nonumber\\ &&
\left(\frac{361179}{4}g_{2}^4-\frac{5273625}{4}g_{2}g_{3}^2\right)\lambda
\nonumber \\&&-972405g_{3}g_{2}^3-1500625g_{3}^3.\label{lame4}
\end{eqnarray}
Using the general formulae, we will consider below the physically
important cases of $n=3,4$ \cite{psv99} which are associated with
the three-gap  $12\wp(\zeta+\omega')$, and four-gap elliptic
potentials $20\wp(\zeta+\omega')$.

The Hermite polynomial ${\mathcal F}(\wp(x),\lambda)$ associated
to the Lam\'e potential $12\wp(\zeta)$ has the form
\begin{eqnarray}
{\mathcal F}(\wp(\zeta),\lambda)=&&\lambda^{3}-
6\wp(\zeta+\omega')\lambda^2 -3\cdot
5(-3\wp(\zeta+\omega')^2+g_{2})\lambda
\nonumber \\
&&-\frac{3^2\cdot
5^2}{4}(4\wp(\zeta+\omega')^3-g_2\wp(\zeta+\omega')-g_{3}).
\label{HerPol3} \end{eqnarray} The solution is real under the
choice of the arbitrary constants $a_{i}, i=1,\ldots ,n$ in such
way, that the constants $a_{i}-\Delta, i=1,\ldots, n$  lie in {\it
different } lacunae. Comparing (\ref{Hansatz}) and (\ref{HerPol3})
and using (\ref{answern1}) the solutions of polynomial equations
(\ref{Peq1}),(\ref{Peq2}),(\ref{Peq3}) can be given by
\begin{eqnarray}
&& A_{i}=\frac{2\cdot 5^2\cdot 3^2}{\prod_{k\neq i}^{3} (a_{i}-a_{k})} , \\
&& B_{i}=-\frac{2\cdot 3^2\cdot 5(a_{i}-\Delta )}{\prod_{k\neq
i}^{n} (a_{i}-a_{k})},\\ && \Delta=\frac{2}{5}\sum_{i=1}^{3}
a_{i}.
\end{eqnarray}
The Hermite polynomial ${\mathcal F}(\wp(\zeta),\lambda)$
associated to the Lam\'e potential $20\wp(\zeta)$ can be written
as
\begin{eqnarray}
{\mathcal
F}(\wp(\zeta),\lambda)&=&11025\wp(\zeta+\omega')^4-1575\wp(\zeta+\omega')^3
\lambda + \nonumber\\&&
(135\lambda^2-\frac{6615}{2}g_{2})\wp(\zeta+\omega')^2+\nonumber
\\&&(-10\lambda^3+ \frac{1855}{4}\lambda g_{2}-2450 g_{3})\wp(\zeta+\omega')+
\nonumber \\&& \lambda^4-\frac{113}{2}\lambda^2 g_{2}+
\frac{3969}{16}g_{2}^2+\frac{195}{4}\lambda g_{3}. \label{HerPol4}
\end{eqnarray}
Comparing (\ref{Hansatz1}) and (\ref{HerPol4}) and using
(\ref{answern1}) the solutions of polynomial equations
(\ref{Pequ1}-\ref{Pequ4}) can be given by
\begin{eqnarray}
&&A_{i}=\frac{11025\cdot 2}{\prod_{k\neq i}(a_{i}-a_{k})},\nonumber \\
&&B_{i}=-\frac{1575\cdot 2(a_{i}-\Delta)} {\prod_{k\neq
i}(a_{i}-a_{k})}\\ && \Delta=\frac{2}{7}\sum_{i=1}^{4}
a_{i}.\nonumber \end{eqnarray} Next solution of system
(\ref{systemn},$n=3$) we obtain using the following ansatz
\begin{eqnarray}
q_{i}(\zeta)=\sqrt{A_{i}\wp(\zeta+\omega')^2+B_{i}\wp(\zeta+\omega')+C_{i}}
 , \quad i=1,2,3,\label{Hansatz-2}
\end{eqnarray}
then we have
\begin{eqnarray}
&&\sum_{i=1}^{3}A_{i}=0,\quad \sum_{i=1}^{3}B_{i}=-6,
\\&& a_{i}=\sum_{k=1}^{3}C_{k}-3\frac{B_{i}}{A_{i}},\quad
C_{i}=\frac{B_{i}^2}{A_{i}}-\frac{1}{4}A_{i} g_{2},
\\&& \frac{{\mathcal C}_{i}^2\cdot 3^3\cdot 4}{A_{i}^2}=
(4\lambda^5+27\lambda^2 g_{3}+ 27\lambda g_{2}^2-21\lambda^3
g_{2}-81 g_{2}g_{3}),
\end{eqnarray}
where $\lambda=-3 B_{i}/A_{i}$. More complicated solution we can
obtain as a special solution of last polynomial system
\begin{eqnarray}
&&q_{1}=C_{1}\alpha\left(\frac{1}{3} C_{-}-k^2\mbox{sn}^2(\alpha
z,k)\right),
\nonumber\\
&&q_{2}=C_{2}\alpha k\mbox{sn}(\alpha z,k)\mbox{cn}(\alpha z,k), \nonumber\\
&&q_{3}=C_{3}\alpha k\mbox{cn}(\alpha z,k)\mbox{dn}(\alpha z,k),
\nonumber
\end{eqnarray}
where
\begin{eqnarray}
&&C_{\pm}=1+k^2\pm\sqrt{1-k^2+k^4}, \nonumber \\
&& k^2=\frac{1}{3}\left(2-\gamma+2\sqrt{\gamma^2-\gamma-2}\right),
\nonumber \\
&& \gamma=\frac{2 a_{3}+a_{2}-3a_{1}}{a_{3}-a_{2}},
\nonumber \\
&&\alpha^2=\frac{1}{3}(a_{3}-a_{2}),\quad
C_{3}^{2}=(k^2-\frac{2}{3}C_{-})C_{1}^2+6, \quad  \nonumber \\
&&C_{2}^2=k^2(C_{1}^2+C_{3}^2) . \nonumber
\end{eqnarray}
Another solution is written by
\begin{eqnarray}
&&q_{1}=C_{1}\alpha\left(\frac{1}{3} C_{-}-k^2\mbox{sn}^2(\alpha
z,k)\right),
\nonumber\\
&&q_{2}=C_{2}\alpha k\mbox{sn}(\alpha z,k)\mbox{cn}(\alpha z,k), \nonumber\\
&&q_{3}=C_{3}\alpha\left(\frac{1}{3} C_{+}-k^2\mbox{sn}^2(\alpha
z,k)\right),
\end{eqnarray}
where
\begin{eqnarray}
&&C_{\pm}=1+k^2 \pm\sqrt{1-k^2+k^4}, \nonumber \\
&&\alpha^2=\frac{(a_{3}-a_{2})}{ 4 + k^2-2 C_{-} },\quad
C_{3}^{2}=-\frac{\left(6+(k^2-\frac{2}{3}C_{-})C_{1}^2\right)}
{k^2-\frac{2}{3} C_{+} },
\quad  \nonumber \\
&&C_{2}^2=k^2(C_{1}^2+C_{3}^2) . \nonumber
\end{eqnarray}

Next solution is given by
\begin{eqnarray}
&&q_{1}=C_{1}\alpha\left(\frac{1}{3} C_{-}-k^2\mbox{sn}^2(\alpha
z,k)\right),
\nonumber\\
&&q_{2}=C_{2}\alpha k\mbox{sn}(\alpha z,k)\mbox{dn}(\alpha z,k), \nonumber\\
&&q_{3}=C_{3}\alpha\left(\frac{1}{3} C_{+}-k^2\mbox{sn}^2(\alpha
z,k)\right),
\end{eqnarray}
where
\begin{eqnarray}
&&C_{\pm}=1+k^2 \pm\sqrt{1-k^2+k^4}, \nonumber \\
&&\alpha^2=\frac{(a_{3}-a_{2})}{ 1 + 4 k^2-2 C_{-} },\quad
C_{3}^{2}=-\frac{\left(6+(1-\frac{2}{3}C_{-})C_{1}^2\right)}
{1-\frac{2}{3} C_{+} },
\quad  \nonumber \\
&&C_{2}^2=C_{1}^2+C_{3}^2 . \nonumber
\end{eqnarray}
From last three solutions in the limit $k\rightarrow 1$ we obtain
the following soliton solutions
\begin{eqnarray}
Q_{1}&=&\sqrt{\frac{9a}{4}} \epsilon_{1}
\exp\left\{i\left(\frac{1}{2}c(x-x_{0})+(a-\frac{1}{4}
c^{2})t\right) \right\}  \times \nonumber \\ &&
\left(-\frac{2}{3}+\mbox{sech}^2(\alpha(x-x_{0}-ct)\right),
\nonumber \\
Q_{2}&=&\sqrt{3(2\alpha^2-a)} \epsilon_{2}
\exp\left\{i\left(\frac{1}{2}c(x-x_{0})+(\alpha^2+a-\frac{1}{4}
c^{2})t\right) \right\}  \times \nonumber \\ && \mbox{th}(\alpha
(x-x_{0}-ct))\mbox{sech}(\alpha(x-x_{0}-ct))
,\nonumber \\
Q_{3}&=&\frac{ \sqrt{3(2\alpha^2-\frac{a}{4}} \epsilon_{3}
\exp\left\{i\left(\frac{1}{2}c(x-x_{0})+(4\alpha^2+a-\frac{1}{4}
c^{2})t\right) \right\} } {\mbox{ch}^{2}(\alpha (x-x_{0}-ct))
},\nonumber
\end{eqnarray}
where we introduce the following notations
\begin{eqnarray}
|\epsilon_{1}|^2=|\epsilon_{2}|^2=|\epsilon_{3}|^2=1, \quad
a_{1}=a,\,a_{2}=\alpha^2+a,\, a_{3}=4\alpha^2+a. \nonumber
\end{eqnarray}

Another type solution in the limit $k\rightarrow 1$ have the
following form
\begin{eqnarray}
Q_{1}&=&C_{1} \epsilon_{1}
\exp\left\{i\left(\frac{1}{2}c(x-x_{0})-(a+\frac{1}{4}
c^{2})t\right) \right\}  \times \nonumber \\ &&
\left(-\frac{2}{3}+\mbox{sech}^2(\sqrt{\frac{a}{8}}(x-x_{0}-ct)\right),
\nonumber \\
Q_{2}&=&C_{2} \epsilon_{2}
\exp\left\{i\left(\frac{1}{2}c(x-x_{0})-(\frac{7}{8}a+\frac{1}{4}
c^{2})t\right) \right\}  \times \nonumber \\ &&
\mbox{th}(\sqrt{\frac{a}{8}}
(x-x_{0}-ct))\mbox{sech}(\sqrt{\frac{a}{8}}(x-x_{0}-ct))
,\nonumber \\
Q_{3}&=&\frac{ \sqrt{\frac{9}{4}a} \epsilon_{3}
\exp\left\{i\left(\frac{1}{2}c(x-x_{0})+(\frac{7}{8}a-\frac{1}{4}
c^{2})t\right) \right\} } {\mbox{ch}^{2}(\sqrt{\frac{a}{8}}
(x-x_{0}-ct)) },\nonumber
\end{eqnarray}
where we introduce the following notations
\begin{eqnarray}
&&|\epsilon_{1}|^2=1,|\epsilon_{2}|^2+|\epsilon_{3}|^2=1, \quad
\nonumber \\
&&a_{1}=-a,\,a_{2}=a_{3}=-\frac{7}{8}a,\,C_{2}^2=C_{1}^2+C_{3}^2.
\nonumber
\end{eqnarray}

\section{Lam\'e polynomials and multicomponent cnoidal waves}
Next we skip the details of derivation of exact solutions and give
only the final results for special case $n=3$ and all $C_j=0$
using methods presented in \cite{eek00}. The $(2n+1)$ Lam\'e
polynomials of order $n$ are solutions of
\begin{eqnarray}
\frac{d^2 E_{i}}{d z^2} + \left(\lambda_{i}
-n(n+1)k^2\mbox{sn}^2(\alpha z) \right) E_{i}=0.
\end{eqnarray}
For $n=3$ we introduce the following eigenfunctions functions
$\mbox{E}_{i},\,i=1,\ldots 7$ and eigenvalues $\lambda_{i}$ given
in table \ref{tab:2}, $\lambda_1<\lambda_2<\ldots <\lambda_7$ and
the constants $C^{(3)}_{i}$ given in table \ref{tab:3}.
\begin{table}[h]
\caption{\label{tab:2}}
\begin{tabular}{|l|l|l|} \hline
1  & $\mbox{E}_{1}^{(3)}=\mbox{sn}(\alpha z,k)\left(
\mbox{dn}^2(\alpha z,k)+C_{1}^{(3)}\right)$ &
$\lambda_{1}^{(3)}=7-5 k^2-2\sqrt{4-7k^2+4k^4}$ \\
\hline 2 & $\mbox{E}_{2}^{(3)}=\mbox{cn}(\alpha z,k)\left(
\mbox{dn}^2(\alpha z,k)+C_{2}^{(3)}\right)$ & $
\lambda_{2}^{(3)}=7-2 k^2-2\sqrt{4-k^2+k^4} $  \\
\hline 3 & $\mbox{E}_{3}^{(3)}=\mbox{dn}(\alpha z,k)\left(
\mbox{dn}^2(\alpha z,k)+C_{3}^{(3)}\right)$ &
$ \lambda_{3}^{(3)}=5(2-k^2)-2\sqrt{1-k^2+4k^4}$   \\
\hline 4 & $\mbox{E}_{4}^{(3)}=\mbox{sn}(\alpha
z,k)\mbox{cn}(\alpha z,k)\mbox{dn}(\alpha z,k)$ &
$ \lambda_{4}^{(3)}=4(2-k^2) $   \\
\hline 5 & $\mbox{E}_{5}^{(3)}=\mbox{sn}(\alpha z,k)\left(
\mbox{dn}^2(\alpha z,k)+C_{5}^{(3)}\right)$ & $
\lambda_{5}^{(3)}=7-5 k^2+2\sqrt{4-7k^2+4k^4}$   \\
\hline 6 & $\mbox{E}_{6}^{(3)}=\mbox{cn}(\alpha z,k)\left(
\mbox{dn}^2(\alpha z,k)+C_{6}^{(3)}\right)$ & $
\lambda_{6}^{(3)}=7-2 k^2+2\sqrt{4-k^2+k^4} $  \\
\hline 7 & $\mbox{E}_{7}^{(3)}=\mbox{dn}(\alpha z,k)\left(
\mbox{dn}^2(\alpha z,k)+C_{7}^{(3)}\right)$ &
$ \lambda_{7}^{(3)}=5(2-k^2)+2\sqrt{1-k^2+4k^4}$    \\
\hline
\end{tabular}
\end{table}
\begin{center}
\begin{table}[h]
\caption{\label{tab:3}}
\begin{tabular}{|l|l|} \hline
1 &
$C_{1}^{(3)}=\frac{1}{5}(2k^2-3-\sqrt{4-7k^2+4k^4})$ \\
\hline 2& $
C_{2}^{(3)}=\frac{1}{5}(k^2-3-\sqrt{4-k^2+k^4}) $  \\
\hline 3 &
$ C_{3}^{(3)}=\frac{1}{5}(2k^2-4-\sqrt{1-k^2+4k^4})$   \\
\hline 4 &
   \\
\hline 5 &  $
C_{5}^{(3)}=\frac{1}{5}(2k^2-3+\sqrt{4-7k^2+4k^4})$   \\
\hline 6 & $
C_{6}^{(3)}=\frac{1}{5}(k^2-3+\sqrt{4-k^2+k^4}) $  \\
\hline 7 &
$ C_{7}^{(3)}=\frac{1}{5}(2k^2-4+\sqrt{1-k^2+4k^4})$    \\
\hline
\end{tabular}
\end{table}
\end{center}
Next we enumerate the periodic solutions of the system
(\ref{systemn}) $\sigma =1$, $s=1$, $n=3$. The results are
collected in Table \ref{tab:4}. For convenience we present
solutions in the following form
\begin{eqnarray}
q_{i}=C_{i}\,\alpha\,\mbox{E}_{i}^{(3)},\quad q_{j}=C_{2}\,\alpha
k^2\,\mbox{E}^{(3)}_{j},\quad q_{k}=C_{k}\,\alpha k\,
\mbox{E}_{k}^{(3)},
\end{eqnarray}
and
\begin{eqnarray}
C_{i}^2=-\Lambda/\Lambda_{i;j,k},\quad
C_{j}^2=\Lambda/\Lambda_{j;i,k},\quad
C_{k}^2=-\Lambda/\Lambda_{k;i,j},
\end{eqnarray}
where $i\neq j \neq k$ and
\begin{eqnarray}
\Lambda=2\cdot 3^2\cdot 5^2,\quad \Lambda_{i;jk}=
(\lambda_{i}^{(3)}-\lambda_{j}^{(3)})
(\lambda_{i}^{(3)}-\lambda_{k}^{(3)}). \nonumber
\end{eqnarray}

\begin{table}[h]
\caption{\label{tab:4}}
\begin{tabular}{|l|l|l|} \hline
A1  & $q_{1}=C_{1}\,\alpha\,\mbox{E}_{3}^{(3)},\quad
q_{2}=C_{2}\,\alpha k^2\,\mbox{E}^{(3)}_{4},\quad
q_{3}=C_{3}\,\alpha k\, \mbox{E}_{6}^{(3)}$ &
$\left\{3,4,6 \right\}$ \\
\hline & $C_{1}^2=-\Lambda/\Lambda_{3;4,6},\quad
C_{2}^2=\Lambda/\Lambda_{4;3,6},\quad
C_{3}^2=-\Lambda/\Lambda_{6;3,4}$ & $
 $  \\
 \hline
A2  & $q_{1}=C_{1}\,\alpha\,\mbox{E}_{3}^{(3)},\quad
q_{2}=C_{2}\,\alpha k\,\mbox{E}^{(3)}_{5},\quad
q_{3}=C_{3}\,\alpha , \mbox{E}_{7}^{(3)}$ &
$\left\{3,5,7 \right\}$ \\
\hline & $C_{1}^2=\Lambda/\Lambda_{3;5,7},\quad
C_{2}^2=-\Lambda/\Lambda_{5;3,7},\quad
C_{3}^2=\Lambda/\Lambda_{7;3,5}$ & $
 $  \\
 \hline
A3  & $q_{1}=C_{1}\,\alpha\,\mbox{E}_{3}^{(3)},\quad
q_{2}=C_{2}\,\alpha k\,\mbox{E}^{(3)}_{5},\quad
q_{3}=C_{3}\,\alpha k\, \mbox{E}_{6}^{(3)}$ &
$\left\{3,5,6 \right\}$ \\
\hline & $C_{1}^2=\Lambda/\Lambda_{3;5,6},\quad
C_{2}^2=-\Lambda/\Lambda_{5;3,6},\quad
C_{3}^2=\Lambda/\Lambda_{6;3,5}$ & $
 $  \\
\hline A4  & $q_{1}=C_{1}\,\alpha\,\mbox{E}_{3}^{(3)},\quad
q_{2}=C_{2}\,\alpha k^2\,\mbox{E}^{(3)}_{4},\quad
q_{3}=C_{3}\,\alpha k\, \mbox{E}_{7}^{(3)}$ &
$\left\{3,4,7 \right\}$ \\
\hline & $C_{1}^2=\Lambda/\Lambda_{3;4,7},\quad
C_{2}^2=-\Lambda/\Lambda_{4;3,7},\quad
C_{3}^2=\Lambda/\Lambda_{7;3,4}$ & $
 $  \\
 \hline
A5  & $q_{1}=C_{1}\,\alpha\,k\mbox{E}_{2}^{(3)},\quad
q_{2}=C_{2}\,\alpha k\,\mbox{E}^{(3)}_{5},\quad
q_{3}=C_{3}\,\alpha k\, \mbox{E}_{6}^{(3)}$ &
$\left\{2,5,6 \right\}$ \\
\hline & $C_{1}^2=\Lambda/\Lambda_{2;5,6},\quad
C_{2}^2=-\Lambda/\Lambda_{5;2,6},\quad
C_{3}^2=\Lambda/\Lambda_{6;2,5}$ & $
 $  \\
 \hline
A6  & $q_{1}=C_{1}\,\alpha\,k\mbox{E}_{2}^{(3)},\quad
q_{2}=C_{2}\,\alpha k\,\mbox{E}^{(3)}_{5},\quad
q_{3}=C_{3}\,\alpha\, \mbox{E}_{7}^{(3)}$ &
$\left\{2,5,7 \right\}$ \\
\hline & $C_{1}^2=\Lambda/\Lambda_{2;5,7},\quad
C_{2}^2=-\Lambda/\Lambda_{5;2,7},\quad
C_{3}^2=\Lambda/\Lambda_{7;2,5}$ & $
 $  \\
 \hline
A7  & $q_{1}=C_{1}\,\alpha\,k\mbox{E}_{2}^{(3)},\quad
q_{2}=C_{2}\,\alpha k^2\,\mbox{E}^{(3)}_{4},\quad
q_{3}=C_{3}\,\alpha \, \mbox{E}_{7}^{(3)}$ &
$\left\{2,4,7 \right\}$ \\
\hline & $C_{1}^2=\Lambda/\Lambda_{2;4,7},\quad
C_{2}^2=-\Lambda/\Lambda_{4;2,7},\quad
C_{3}^2=\Lambda/\Lambda_{7;2,4}$ &
 $ $ \\
 \hline
A8  & $q_{1}=C_{1}\,\alpha\,k\mbox{E}_{2}^{(3)},\quad
q_{2}=C_{2}\,\alpha k^2\,\mbox{E}^{(3)}_{4},\quad
q_{3}=C_{3}\,\alpha \, \mbox{E}_{6}^{(3)}$ &
$\left\{2,4,6 \right\}$ \\
\hline & $C_{1}^2=\Lambda/\Lambda_{2;4,6},\quad
C_{2}^2=-\Lambda/\Lambda_{4;2,6},\quad
C_{3}^2=\Lambda/\Lambda_{6;2,4}$ & $
 $  \\
\hline
\end{tabular}
\end{table}
For defocusing case $\sigma=-1$, $s=1$, $n=3$ we have presented
solutions in Table \ref{tab:5}.

\begin{table}[h]
\caption{\label{tab:5}}
\begin{tabular}{|l|l|l|} \hline
B1  & $q_{1}=C_{1}\,\alpha\,k\mbox{E}_{1}^{(3)},\quad
q_{2}=C_{2}\,\alpha \,\mbox{E}^{(3)}_{3},\quad q_{3}=C_{3}\,\alpha
k\, \mbox{E}_{5}^{(3)}$ &$\left\{1,3,5 \right\}
 $
 \\
\hline & $C_{1}^2=-\Lambda/\Lambda_{1;3,5},\quad
C_{2}^2=-\Lambda/\Lambda_{3;1,5},\quad
C_{3}^2=\Lambda/\Lambda_{5;1,3}$ & $
 $  \\
 \hline
B2  & $q_{1}=C_{1}\,\alpha\,k\mbox{E}_{1}^{(3)},\quad
q_{2}=C_{2}\,\alpha k\,\mbox{E}^{(3)}_{2},\quad
q_{3}=C_{3}\,\alpha ,k \mbox{E}_{5}^{(3)}$ &
$\left\{1,2,5 \right\}$ \\
\hline & $C_{1}^2=\Lambda/\Lambda_{1;2,5},\quad
C_{2}^2=-\Lambda/\Lambda_{2;1,5},\quad
C_{3}^2=\Lambda/\Lambda_{5;1,2}$ & $
 $  \\
 \hline
B3  & $q_{1}=C_{1}\,\alpha\,k\mbox{E}_{1}^{(3)},\quad
q_{2}=C_{2}\,\alpha k\,\mbox{E}^{(3)}_{2},\quad
q_{3}=C_{3}\,\alpha \, \mbox{E}_{4}^{(3)}$ &
$\left\{1,2,4 \right\}$ \\
\hline & $C_{1}^2=\Lambda/\Lambda_{1;2,4},\quad
C_{2}^2=-\Lambda/\Lambda_{2;1,4},\quad
C_{3}^2=\Lambda/\Lambda_{4;1,6}$ & $
 $  \\
\hline B4  & $q_{1}=C_{1}\,\alpha\,k\mbox{E}_{1}^{(3)},\quad
q_{2}=C_{2}\,\alpha \,\mbox{E}^{(3)}_{3},\quad q_{3}=C_{3}\,\alpha
k^2\, \mbox{E}_{4}^{(3)}$ &
$\left\{1,3,4 \right\}$ \\
\hline & $C_{1}^2=\Lambda/\Lambda_{1;3,4},\quad
C_{2}^2=-\Lambda/\Lambda_{3;1,4},\quad
C_{3}^2=\Lambda/\Lambda_{4;1,3}$ & $
 $  \\
\hline
\end{tabular}
\end{table}

{\bf Acknowledgements.} This work has been supported  by the
National Science Foundation of Bulgaria, contract No. F-1410.

\section{Conclusions}

In this paper we find five different families of periodic
solutions of optical cascading equations related to eigenvalues of
Lam\'{e} equation with two-gap potential. We investigate also
localized solitary wave solutions as a special limit of periodic
solutions.

The existence of two-gap solutions expressed as product of two
elliptic functions can be viewed as manifestation of "partial
integrability" of these equations. We expected that both group
theoretical method \cite{lwm98} and spectral method will lead us
to new understanding of "partial integrability" of optical
cascading equations.

For vector nonlinear Schr\"{o}dinger equation $n=3,4$ we present
exact solutions in a form of multicomponent cnoidal waves.
Multicomponent cnoidal waves are previously discussed in
\cite{psv99}.

{\bf Acknowledgements.} This work has been supported  by the
National Science Foundation of Bulgaria, contract No. F-1410.

\end{document}